\documentclass[12pt]{article}

\usepackage[numbers,square,sort&compress]{natbib}
\usepackage{url}
\usepackage{breakurl}

\usepackage[breaklinks]{hyperref}
\usepackage{xcolor}
\usepackage[margin=2.5cm]{geometry}
\hypersetup{
    colorlinks,
    linkcolor={red!50!black},
    citecolor={blue!50!black},
    urlcolor={blue!80!black}
}
\usepackage{graphicx}
\usepackage{amsmath}
\usepackage{mathtools}

\newcommand{\fig}[3][1]{
    \begin{figure}[h]
        \centering
        \includegraphics[width=#1\textwidth]{fig/#2}
        \caption{#3 \label{fig/#2}}
    \end{figure}
    }

\newcommand{\figref}[2]{Figure\,\ref{fig/#1}{#2}}

\newcommand{\secref}[1]{Section \ref{sec/#1}}

\newcommand{\appref}[1]{Appendix \ref{app/#1}}

\newcommand{\tab}[5]{ 
    \begin{table}     
        \centering
        \caption{
            #2 \label{tab/#1}
        }    
        \begin{tabular}{#3} 
        \hline 
            #4 \\
        \hline
         #5 
        \end{tabular}
    \end{table}
}

\newcommand{\tabref}[1]{Table\,\ref{tab/#1}}

\newcommand{\eq}[2]{
    \begin{equation}
        #2 \label{#1}
    \end{equation}
}

\newcommand{\eqarr}[1]{
    \begin{eqnarray}
        #1
    \end{eqnarray}
}

\DeclareMathOperator*{\argmax}{arg\,max}
\DeclareMathOperator*{\argmin}{arg\,min}

\renewcommand\Re{\mathrm{Re}\,} 
\renewcommand\Im{\mathrm{Im}\,} 

\DeclareMathOperator{\atantwo}{atan2}

\newcommand{\ga}{\ensuremath{\gamma}}
\newcommand{\GA}{\ensuremath{\Gamma}}
\newcommand{\de}{\ensuremath{\delta}}

\newcommand{\ep}{\ensuremath{\epsilon}}

\newcommand{\et}{\ensuremath{\eta}}

\renewcommand{\th}{\ensuremath{\theta}}

\newcommand{\la}{\ensuremath{\lambda}}
\newcommand{\LA}{\ensuremath{\Lambda}}

\newcommand{\si}{\ensuremath{\sigma}}
\newcommand{\SI}{\ensuremath{\Sigma}}
\newcommand{\ta}{\ensuremath{\tau}}

\newcommand{\ph}{\ensuremath{\phi}}
\newcommand{\PH}{\ensuremath{\Phi}}
\newcommand{\om}{\ensuremath{\omega}}

\newcommand{\PO}{\mathrm{PO}}
\newcommand{\inprod}[2]{\ensuremath{\langle #1, \, #2 \rangle}}

\newcommand{\oF}{\ensuremath{\mathcal{F}}}
\newcommand{\oJ}{\ensuremath{\mathcal{J}}}
\newcommand{\oL}{\ensuremath{\mathcal{L}}}
\newcommand{\oM}{\ensuremath{\mathcal{M}}}

\newcommand{\oR}{\ensuremath{\mathcal{R}}}

\newcommand{\oI}{\ensuremath{\mathcal{I}}}
\newcommand{\oT}{\ensuremath{\mathcal{T}}}

\newcommand{\reals}{\mathbb{R}}
\newcommand{\KE}{\mathrm{KE}}
\usepackage{amsfonts} 
\usepackage{cleveref}
\usepackage{subcaption}    
\usepackage{comment}

\title{ 
    Prediction and control of spatiotemporal chaos by \emph{learning} conjugate
    tubular neighborhoods 
    }

\author{
    Nazmi Burak Budanur  \\ 
    \small{Max Planck Institute for the Physics of Complex Systems (MPIPKS)}\\
    \small{Nöthnitzer Straße 38, 01187 Dresden, Germany}
    }

\date{\today}

\begin{document}

\maketitle

\begin{abstract}
    I present a data-driven predictive modeling tool that is applicable to
    high-dimensional chaotic systems with unstable periodic orbits. The basic
    idea is using deep neural networks to \emph{learn} coordinate
    transformations between the trajectories in the periodic orbits'
    neighborhoods and those of low-dimensional linear systems in \emph{a latent
    space}. I argue that the resulting models are partially \emph{interpretable}
    since their latent-space dynamics is fully understood. To illustrate the
    method, I apply it to the numerical solutions of the
    Kuramoto--Sivashinsky partial differential equation in one dimension.
    Besides the forward-time predictions, I also show that these models can be
    leveraged for control. 
\end{abstract}

\section{Introduction}
\label{sec/Introduction}

The trade-off between interpretability and accuracy is encountered in many
facets of data science \cite{samek2019explainable,miller2019explanation}.
Typically, the more complicated a model is, the higher is its prediction
accuracy; and conversely, human-interpretable models perform poorly in complex
tasks. In the data-driven studies of dynamical systems, this conundrum presents
itself as unexplainable (black-box) models of chaos, especially in systems with
many degrees of freedom, e.g. spatiotemporal phenomena
\cite{pathak2018modelfree,vlachas2018datadriven}. While such black-box models
could be immensely useful for making predictions, especially in systems without
known models, their contribution to the basic understanding of a system is
little. 

While several interpretable data-driven dynamical modeling tools exist in the
literature, their applications have been predominantly limited to
low-dimensional chaos or high-dimensional non-chaotic systems.
Arguably the most popular interpretable dynamical system modeling method is the
dynamic mode decomposition (DMD)
\cite{schmid2010dynamic,kutz2016dynamic,schmid2022dynamic}, with which one seeks
a best-fit linear system to a given dataset and perform a modal expansion for
predictions. Although a rigorous connection between DMD modes and the
eigenfunctions of the Koopman operator exists \cite{rowley2009spectral}, in
practice the conditions for this correspondence are rarely satisfied except in
the close neighborhoods of fixed points \cite{page2019koopman} and periodic
solutions \cite{rowley2009spectral}. Alternative approaches
\cite{lusch2018deep,otto2019linearly} to the Koopman formalism utilized
deep learning for identifying the Koopman eigenfunctions, which evolve 
linearly. While the applications of Ref. \cite{lusch2018deep} included 
non-chaotic nonlinear systems such as nonlinear pendulum and wake flow behind 
a cylindrical obstacle, Ref. \cite{otto2019linearly} demonstrated that such 
models can also predict the evolution of the spatiotemporally chaotic 
Kuramoto--Sivashinsky system for short times. Another recent study \cite{cenedese2022datadriven}
showed that one can develop reduced-order models of non-linearizable dynamics in
terms of spectral submanifolds, which are smooth extensions of the eigenspaces
of invariant solutions, e.g. equilibria and (quasi-)periodic orbits. Once again,
the applications that were considered in \cite{cenedese2022datadriven} were
systems that did not exhibit chaos. For low-dimensional chaotic systems, sparse
identification methods \cite{brunton2016discovering} were shown to be able to
identify the governing equations as well as the intrinsic coordinates on which
the dynamics is sparse \cite{champion2019datadriven}. These methods, however,
rely on a set of candidate terms which increase in number very quickly
with system dimension, thus, are not appropriate for systems with
high-dimensional attractors. 

One of the defining features of chaotic systems is
the presence of a dense set of unstable periodic orbits
\cite{guckenheimer1983nonlinear}.
A central motivation for the present work stems from the numerical discoveries
\cite{kawahara2001periodic,viswanath2007recurrent,cvitanovic2010geometry,
chandler2013invariant,budanur2017relative,suri2020capturing,yalniz2021coarse,
crowley2022turbulence} of unstable periodic orbits in two- and three-dimensional
Navier--Stokes simulations. These papers illustrated various similarities
between the weakly turbulent flows and the periodic orbits and suggested a
Markovian description of turbulence as transitions among the neighborhoods of
these solutions in the state space. In addition, several
\cite{viswanath2007recurrent,
chandler2013invariant,budanur2017relative,crowley2022turbulence} of these papers
also reported that the periodic orbits had varying number of unstable
dimensions, which can be taken as a numerical evidence for the lack of
hyperbolicity \cite{kostelich1997unstable}. An important consequence of the
absence of hyperbolicity is the breakdown of \emph{shadowing}
\cite{dawson1994obstructions}, which implies that the long trajectories arising
in Navier--Stokes simulations might not correspond to ``true'' trajectories of
the system as they explore the state space regions with different number of
unstable directions. In light of the above-cited numerical results and their
theoretical implications, I believe that reduced-order modeling of
high-dimensional chaotic systems such as turbulent flows should not aim to model
the system as a whole but rather attempt to produce an ensemble of models each
applying to distinct regions of the state space. 

Reduced-order data driven models of high-dimensional systems that are local in
the state space can be obtained by performing \emph{clustering} prior to
modelling, see \cite{kaiser2014clusterbased,floryan2022datadriven} for examples.
Both of these studies k-means clustering \cite{lloyd1982least}, which groups
states of the system according to their distances (measured in some norm in the
state space, typically $L_2$) to one another. This approach to the state space
partitioning is, however, blind to the dynamics since two nearby trajectories of
a nonlinear system can evolve towards very different future states if, for
example, they lie at two sides of a basin boundary. This can be readily seen in
Ref. \cite{kaiser2014clusterbased}'s application to the Lorenz system where the
trajectories that evolve towards different ``ears'' of the attractor are grouped
into the same cluster. In contrast, the standard binary symbolic partition of
the Lorenz attractor from which one can accurately estimate fractal properties
of the attractor using the periodic orbit theory \cite{cvitanovic1991periodic}
encodes the trajectories according to the order in which they visit each ear,
see e.g. \cite{eckhardt1994periodic,viswanath2004fractal}.

The purpose of the present paper is to demonstrate through an application to the
one-dimensional Kuramoto--Sivashinsky equation that one can produce simple
data-driven models that are capable of predicting spatiotemporally chaotic
dynamics by anchoring the models at the periodic orbits. Similar to
\cite{lusch2018deep} and \cite{otto2019linearly}, I utilize deep
autoencoders to find transformations between the system's states and a
low-dimensional latent space in which the dynamics is linear. The latent space
dynamics is determined to be conjugate to that of the periodic orbit's
leading Floquet eigenspace. In this sense, the method is conceptually
similar to the spectral subspaces of Ref. \cite{cenedese2022datadriven}, although
here I only aim to capture linearizable dynamics. The paper is organized as
follows. \secref{Kuramoto--Sivashinsky_system} presents a brief description of
the Kuramoto--Sivashinsky equation and the computational tools and introduces
the periodic orbits. The modeling method and its training is described in
\secref{Conjugate_tubular_neighborhoods} followed by the main results in
\secref{Results}. Results are discussed in \secref{Discussion} and the paper
concludes in \secref{Conclusion}. 

\section{Kuramoto--Sivashinsky system}
\label{sec/Kuramoto--Sivashinsky_system}
\subsection{Numerical setup}

\fig{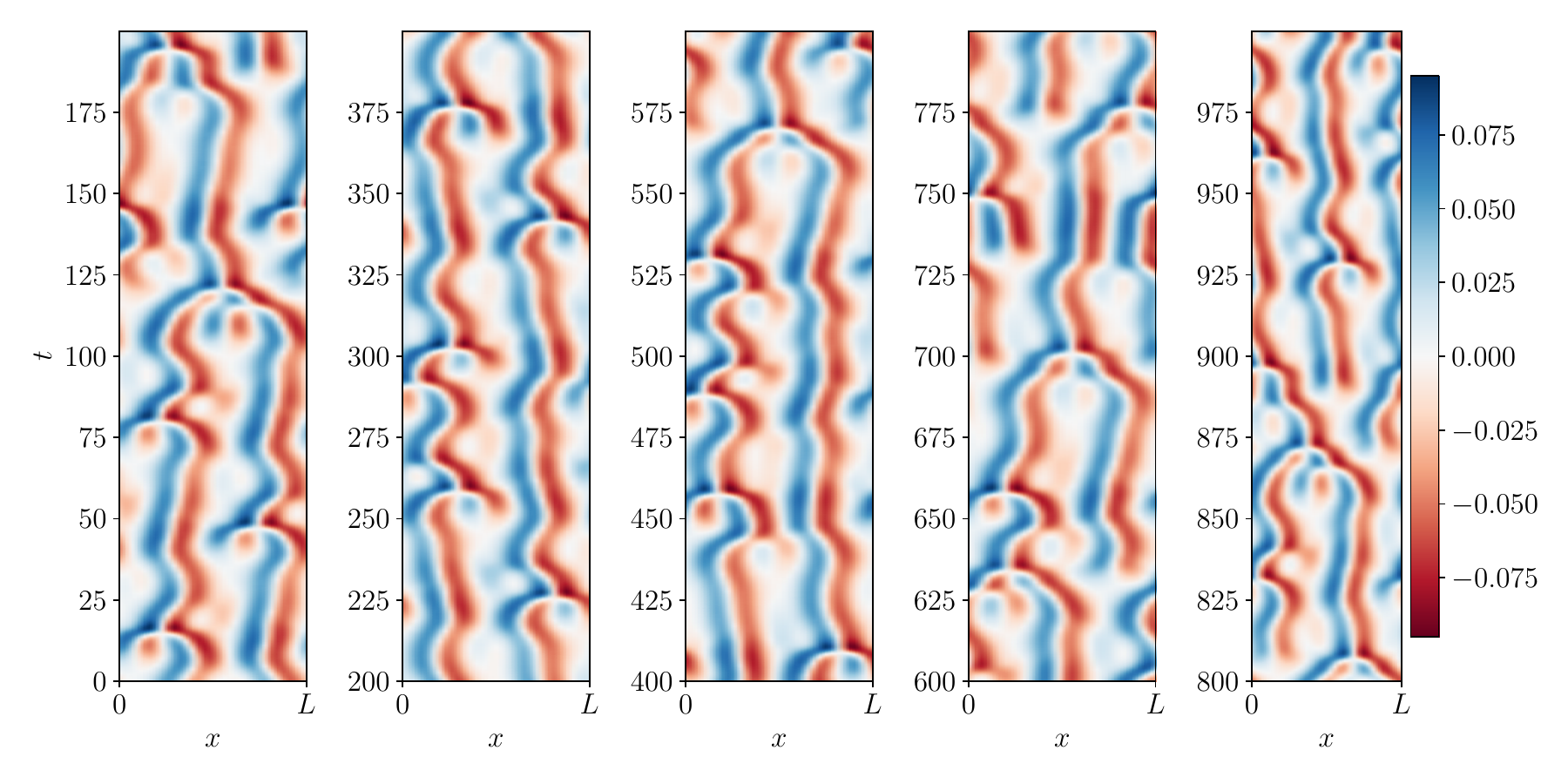}{A spatiotemporally chaotic trajectory visualized as a spacetime
    plot where the amplitude of $u(x,t)$ is color coded. The colorbar shown on
    the right applies to all space-time plots of this paper.}

In one space dimension, the
Kuramoto--Sivashinksy\cite{kuramoto1978diffusioninduced,sivashinsky1977nonlinear}
equation reads 
\eq{KS}{ 
    u_t = - u_{xx} - u_{xxxx} - \frac{1}{2} (u^2)_x 
} 
where the subscripts $_t$ and $_x$ denote the partial derivatives with respect
to time and space, respectively; and the real-valued scalar field $u(x,t)$
satisfies the periodic boundary condition $u(x, t) = u(x + L, t)$. The domain
length $L$ is the sole control parameter of the system and when it is
sufficiently large, the Kuramoto--Sivashinsky system exhibits chaos
\cite{pumir1983on,nicolaenko1985some}. \eqref{KS} can be simulated by computing
the truncated discrete Fourier expansion $u(x) = \sum_{k = -K}^{K} v_k e^{i
q_k x}$, where $q_k = 2 \pi k / L$, and integrating the set of ordinary
differential equations (ODEs)
\eq{FKS}{ 
    \dot{v}_k = (q_k^2 - q_k^4) v_k - \frac{i q_k}{2} 
        \sum_{m =-K}^{K}  v_m v_{k - m} \,, 
} 
which is obtained by substituting $u(x)$ with its Fourier series in \eqref{KS}.
Noting that $k=0$ mode is decoupled from the rest in \eqref{FKS} (Galilean
invariance), one can set it to $0$ without loss of generality to obtain a
$2K$-dimensional dynamical system defined by the nonlinear ODEs \eqref{FKS}. 
Since $v_k = v^*_{-k}$ due to real-valuedness of $u(x,t)$, the number of
independent degrees of freedom of this system is also $d=2K$.
For the numerical integration of \eqref{FKS}, I utilized the general purpose
integrator \texttt{odeint} of \texttt{scipy.integrate} \cite{virtanen2020scipy}
evaluating the nonlinear term pseudospectrally \cite{canuto2007spectral} using
$\sum_{m =-K}^{K}  v_m v_{k - m} = \oF\{u^2\}$, where $\oF$ denotes the 
discrete Fourier transformation. I chose $L=22.0$ and truncated the Fourier
expansion at $K=15$ following Cvitanovi\'c et al. \cite{cvitanovic2010on}, who
reported chaotic dynamics and (relative) periodic orbits at these parameters. In
all results presented in the following, the simulated trajectories are sampled
at time steps of $\de t = 0.01$. As an example, \figref{spacetime}{} shows a
spacetime plot of $u(x,t)$ simulated starting from a random initial condition on
the system's attractor.

When $u(x,t)$ describes a velocity field in one dimension, the $L_2$ inner
product 
\eq{energy}{
    \KE = \frac{1}{2L} \int_0^{L} u^2  dx
    = \frac{1}{2} \sum_{m = -K}^{K} |v_k|^2
}
is interpreted as the kinetic energy density and its rate of change is given by
\cite{cvitanovic2010on}
\eq{dEdt}{
    \dot{\KE} = P - \varepsilon \,,\quad 
    \mbox{where}\quad
    P = \frac{1}{L} \int_0^{L} u_x^2  dx\, \quad \mbox{and}\quad 
    \varepsilon = \frac{1}{L} \int_0^{L} u_{xx}^2  dx\,.
}
One can interpret the observables $P$ and $\varepsilon$ as the instantaneous rates of
power input and dissipation, respectively. 

\subsection{Symmetry reduction}

The Kuramoto--Sivashsinky equation \eqref{KS} in the periodic domain preserve
its form (equivariance) under the translations $u(x, t) \rightarrow u(x - \de x,
t)$ and the reflection $u(x, t) \rightarrow - u (-x, t)$.
Consequently, its chaotic solutions come in families of symmetry copies that can
be generated by these transformations. Additionally, the presence of continuous
symmetries give rise to \emph{relative} periodic orbits, satisfying
\eq{rpo}{
    u_p = \oM^{n_p} \oT_{- \de x_p} \PH^{T_p} (u_p) \,, 
}
where $\de x_p \in [0, L)$, $n_p \in \{0, 1\}$, $\oT_{\de x}$ and $\oM$ are the
translation and reflection (mirror) operators, respectively, and $\PH^{t}$
denotes the finite-time flow implied by the simulation of \eqref{KS}, i.e. $u(t)
= \PH^{t}(u(0))$. In the state space, the relative periodic orbits with
translations correspond to two-tori (quasiperidoicity) parametrized by spatial
and temporal shifts. If $\de x_p = 0$ and $n_p = 1$, then the orbit is said to be
pre-periodic because $u_p$ repeats itself after two periods since $\oM^2 =
\oI$, where $\oI$ is the identity operator. Note that the orbits with $\de x_p
\neq 0$ and $n_p=1$ can be transformed to preperiodic orbits by a translation by
$- \de x_p / 2$ since $\oT_{\de x} \oM = \oM \oT_{-\de x}$. For the model
reduction method presented in the following, these symmetry multiplicities are
undesirable or in some cases prohibitive. Thus, one has to eliminate them by a
symmetry-reduction prior to the dynamical modeling. 

Symmetry reduction of the Kuramoto--Sivashinsky equation by means of first
Fourier mode slice and invariant polynomials was formulated in
\cite{budanur2015reduction,budanur2016unstable}. Already in
\cite{budanur2015reduction} it was shown that eliminating the translation degree
of freedom by fixing the phase of the first Fourier mode to a set value results
in fast oscillations. Although this can be remedied by a time-rescaling
transformation, it may not be possible in all cases, especially if the data is
collected experimentally. Therefore, here I opt for a two-step transformation
starting with one that fixes the phase of the second Fourier mode to a constant
value, followed by construction of complex polynomials that eliminate the
remaining discrete symmetries. Leaving the details to
the \appref{symmred}, hereafter, I use $\xi$ and $S(u)$ to denote the
symmetry-reduced states and the symmetry-reducing transformation satisfying 
\eq{symmetry_reduction}{
    \xi = S(u) = S(\ga u) \,, 
}
where $\ga \in \GA = \{\oT_{\de x}, \oT_{\de x} \oM\}$, such that an inverse
transformation $S^{-1}(\xi) = u' = \ga' u$, where $\xi \in \GA$ can be found. As
a result, no information other than the symmetry multiplicity is lost by the
symmetry reduction since it is revertible up to a choice of a symmetry copy. 

\subsection{The attractor and the periodic orbits}
\label{sec/attractor_and_pos}

\fig{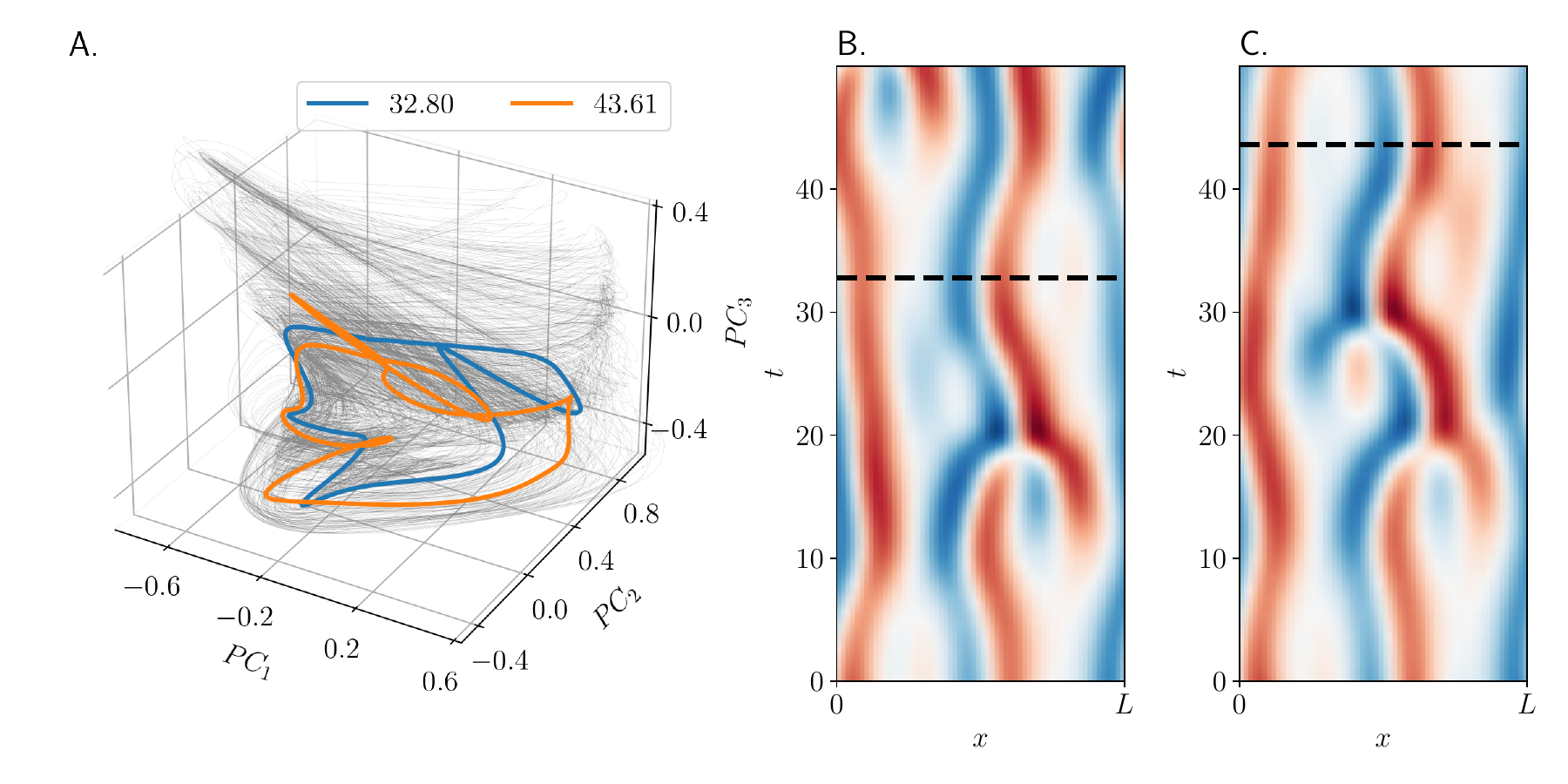}{(A) A chaotic trajectory (gray) spanning a time interval of $t
    \in [0, 10^4]$ along with two periodic orbits $\PO_{1}$ and $\PO_{2}$
    visualized as projections from the $30$-dimensional symmetry-reduced state
    space onto the leading three principal components $PC_{1,2,3}$ of the
    attractor. (B, C) Space-time visualizations of $\PO_{1}$ (B) and $\PO_{2}$
    (C) for a time interval $t \in [0, 50]$. The periods $T_1= 32.80$ and
    $T_2 = 43.61$ of the relative periodic orbits are indicated by the
    horizontal dashed lines in B and C.}

\figref{chaosandpos}{A} illustrates the chaotic attractor of the
Kuramoto--Sivashinsky system as the projection of a symmetry-reduced trajectory
integrated for a time interval $t \in [0, 10^4]$ onto the first three modes 
obtained from the principal
component analysis (PCA) \cite{jolliffe2002principal} of the trajectory. In addition to the
attractor, \figref{chaosandpos}{A} also shows two periodic orbits $\PO_{1}$
and $\PO_{2}$ as blue and orange closed curves, respectively.
\figref{chaosandpos}{B and C} show the spacetime visualizations of
$\PO_{1}$ and $\PO_{2}$, respectively, where each orbit is plotted for
$t \in [0, 50]$ so that adjacent panels have the same timescale and the periods
$T_1= 32.80$ and $T_2 = 43.61$
of the orbits are indicated by the horizontal dashed lines.  While $\PO_{1}$
shifts by $\de x_{1} = 10.96$ after one period, $\PO_{2}$ is
pre-periodic, namely its final state is the reflection of its initial one. These
orbits are the two that I found most-frequently by initiating Newton searches
from close recurrences of the symmetry reduced chaotic trajectories, and they
form the bases of the reduced-order models described in the following. 

\tab{floquet}{
    Leading non-marginal Floquet multipliers and exponents of the periodic 
    orbits.
}{
    c c c c c 
}{
    $\PO$  & $\LA_1$ & $\LA_2$ & $\la_1$ & $\la_2$ 
}{
    $\PO_{1}$ & $0.3166 + i 1.8136$  & $0.3166 - i 1.8136$  
    & $0.0128 + i 0.0293$  & $0.0128 - i 0.0293$ \\
    $\PO_{2}$ & $9.8292$  & $0.0908$  & $0.0480$  & $-0.0504$ 
}

The linear stability of a relative periodic orbit is determined by the Jacobian
of the \eqref{rpo}'s right-hand side given by 
\eq{jacrpo}{
    \oJ_p = \oM^{n_p} \oT_ {- \de x_p} \frac{d \PH^{T_p} (u)}{d u} \Big|_{u = u_p} \,, 
}
whose eigenvalues $\LA_i$ and eigenvectors $V_i$ are called the Floquet
multipliers and Floquet vectors, respectively \cite{cvitanovic2016chaos}.
Corresponding exponents $\la_i$ that satisfy $\LA_i = e^{\la_i T_p}$ are known
as the Floquet exponents and those with negative and positive real parts
correspond to the stable and unstable directions, respectively. Each relative
periodic orbit of the Kuramoto--Sivashinsky system has at least two \emph{marginal}
eigenvalues with $\Re \la_i = 0$ corresponding to directions corresponding to
temporal and spatial translations. \tabref{floquet} lists the leading (ordered
in descending $|\LA_i|$) two non-marginal Floquet multipliers and exponents of
the periodic orbits. These eigenvalues were obtained by approximating the
Jacobian $d\PH^{t}(u)/du$ along the periodic orbit in the Fourier space by
integrating the gradient of \eqref{FKS} starting from the identity matrix as the
initial condition. 

\section{Conjugate tubular neighborhoods}
\label{sec/Conjugate_tubular_neighborhoods}
\fig{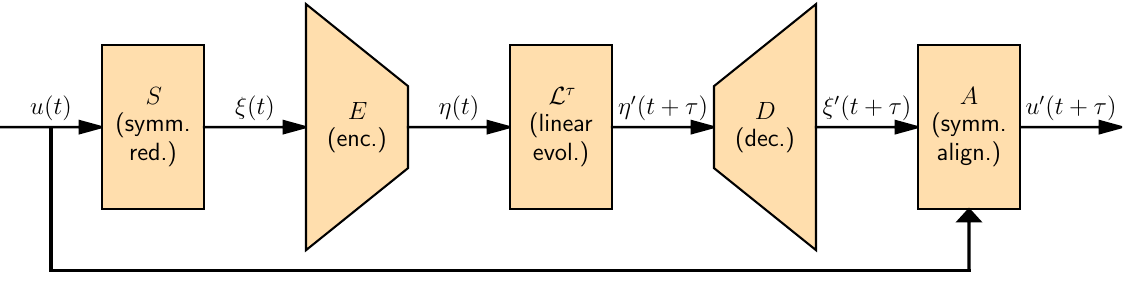}{Block diagram illustrating the proposed model architecture.}
    
\subsection{Basic assumptions and the model architecture}
\label{sec/basic_assumptions}

    I would like to begin with an overview of the modeling approach.
    As \figref{chaosandpos}{A} illustrates, the unstable periodic orbits of the
    Kuramoto--Sivashinsky system appear to be embedded in its chaotic
    attractor. Remembering also that these orbits are identified by Newton searches
    initiated from the states on the attractor, it is sensible to assume that these
    orbits can form the basis of models for approximating the chaotic trajectories
    in their vicinity. 
    Guided by these observations, I construct two models each aiming to predict
    the dynamics in the neighborhood of one of the periodic orbits. The series
    of operations for obtaining predictions from each model is summarized in
    \figref{architecture}{}, where an initial state $u(t)$ is first symmetry
    reduced and then \emph{encoded} into a three-dimensional latent state $\et$.
    The latent state is evolved linearly to $\et'(t + \ta)$, and \emph{decoded}
    back to a symmetry-reduced state $\xi'(t + \ta)$ which is finally
    transformed back to $u'(t + \ta)$ by the inverse symmetry reduction and
    state alignment. Here, $'$ signifies that the quantity is a
    \emph{prediction}. The transformations to and from the latent space are
    achieved by an autoencoder \cite{kramer1991nonlinear} which is trained
    indivdiually for each periodic orbit. The linear dynamics in the latent
    space is prescribed such that the periodic orbit is mapped to the unit
    circle and the transverse dynamics is conjugate
    \cite{guckenheimer1983nonlinear} to that of the \emph{leading} Floquet
    eigenspace of the periodic orbit associated with the exponents with
    the largest real parts. In doing so, the model aims to capture the dynamics
    associated with the exponentially dominant part of the periodic orbits'
    tangent space.

The symmetry reduction $S$ and its inverse $S^{-1}$ is described in the
\secref{Kuramoto--Sivashinsky_system} and \appref{symmred}. The
symmetry-aligning transformation $A(\xi'(t + \tau); u(t))$ shown in
\figref{architecture}{} is an abstraction for the following series of
operations. Given a symmetry-reduced prediction $\xi'(t + \ta)$ for $\ta \in
\{0, \de t, 2 \de t, \ldots \}$, first the inverse symmetry reduction
$S^{-1}(\xi') = \tilde{u}'$ is obtained. Next, for each time step with $\ta>0$, 
one out of four discrete symmetry copies that can be reached by $\si \in \SI =
\{\oI, \oM, \oT (L / 2), \oM \oT\}$  is selected, such that the resulting trajectory is the
smoothest one. This is implemented by approximating the partial derivative
$\tilde{u}'_t$ via finite differences (second-order central differences for
intermediate points and first-order finite-differences at the beginning and the
end of the interval) and transforming $\tilde{u}'(\ta') \rightarrow \si^*(\ta')
\tilde{u}'(\ta')$, where $\ta' = t + \ta$ and $\si^* (\ta') = \argmax_{\si \in
\SI} \inprod{\si \tilde{u}'_t (\ta')}{\tilde{u}'_t (\ta' - \de t)}$. The
resulting trajectory is one where
each state has a fixed second Fourier mode phase, which can be interpreted as a
slice \cite{budanur2015reduction} and, thus, the remaining translations can be
determined by the reconstruction equation \cite{rowley2000reconstruction}.
Finally, $\ga^* = \argmin_{\ga \in \GA} \| u(t) - \ga u'(t) \| $ is found
and applied to the reconstructed trajectory, so that the initial states are
aligned. Note that even though the inverse symmetry reduction $S^{-1}$ is
exact, the selection of the discrete symmetry copy relies on the  
estimation of $\tilde{u}'_t$, which is a source of numerical errors. It is,
therefore, important that the symmetry-reducing transformation does not result
in a loss of temporal resolution.
    
In ref. \cite{bramburger2021deep}, which was influential for the present one,  
Bramburger et al. showed that deep neural networks can be utilized for
discovering conjugacies between discrete-time dynamical systems, i.e. mappings.
Although in principle the approach presented in \cite{bramburger2021deep} can be
extended to continuous-time systems by means of Poincar\'e sections, in practice  
finding suitable Poincar\'e sections become very challenging in high-dimensional
systems because the transversality between the dynamics and a Poincar\'e section
hyperplane is not guaranteed beyond a close neighborhood of a periodic orbit.
For this reason, in this work I formulate a conjugacy-based predictive modeling
tool directly for the continuous-time system.

\fig{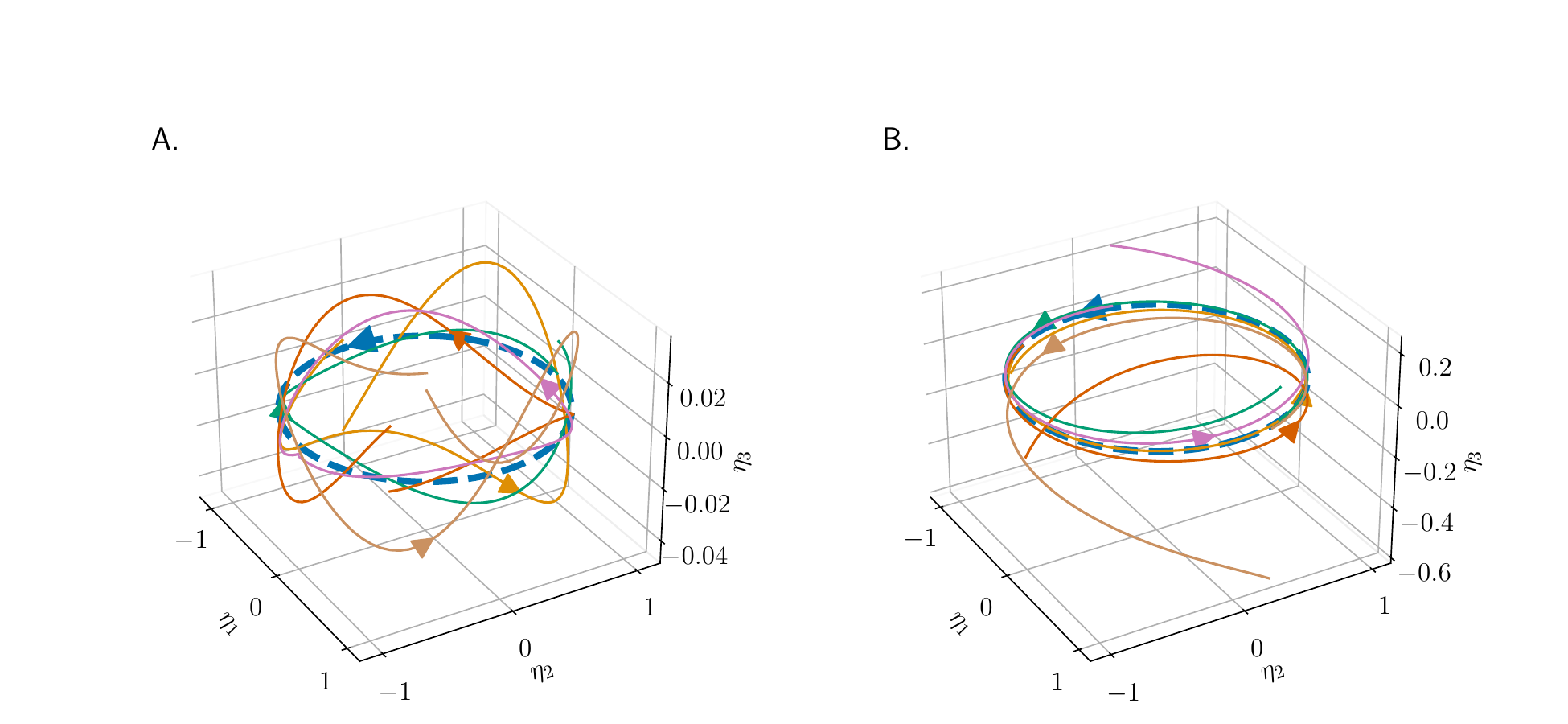}{ Trajectories of the linear systems \eqref{linear_cc} with $(\mu,
    \om, T_p) = (0.2, 2.2, 2\pi)$ (A) and \eqref{linear_rr} with $(\mu_1, \mu_2,
    T_p) = (0.6, -0.1, 2\pi)$ (B). Unit circles corresponding to the periodic
    orbits are shown dashed. Arrowheads along the trajectories indicate the
    direction of time. }

The encoder-decoder pair $(E, D)$ constitutes an autoencoder
\cite{kramer1991nonlinear} which I approximate using neural networks from
$30$-dimensional symmetry-reduced state space to $3$-dimensional latent space
and backwards. Before describing the training through which the network
parameters are determined, I must explain the linear time evolution in the
latent space, which corresponds to the innermost block in \figref{architecture}.
Let $\et = (\et_1, \et_2, \et_3)$ and $R_i (\th)$ be the $3\times3$ rotation
matrix, action of which rotates $\et$ about $\et_i$ by $\th$. If the leading
Floquet exponents of the periodic orbit with the period $T_p$ are a complex
conjugate pair $\la_{1,2} = \mu \pm i \om $, the corresponding latent-space
evolution is given by 
\eq{linear_cc}{ 
    \oL^{\ta} \et = R_3 (\th_0 +  2 \pi \tau / T_p) 
        \left\{  e^{\mu \ta} R_2(\om \ta) 
        \left[R_3(-\th_0) \et - \hat{\et}_1\right] + \hat{\et}_1 \right\} \,, 
    }
where $\th_0 = \atantwo(\et_2, \et_1)$ and $\hat{\et}_1 = (1,0,0)$. The second
case is the purely real Floquet exponents $\mu_{1,2}$ with the
latent-space dynamics given by 
\eq{linear_rr}{ 
    \oL^{\ta} \et = R_3 (\th_0 +  2 \pi \tau / T_p) 
        \left\{
            \left[\hat{\et}_3 e^{\mu_1 \ta} \hat{\et}_3 
            + \hat{\et}_1 e^{\mu_2 \ta} \hat{\et}_1 \right] \cdot 
        \left[R_3(-\th_0) \et - \hat{\et}_1\right] + \hat{\et}_1 \right\} \,, 
    }
where $\hat{\et}_3 = (0,0,1)$. Note that the term
$R_3(-\th_0) \et - \hat{\et}_1$ that appears in both \eqref{linear_rr} and
\eqref{linear_cc} measures $\et$'s deviation from the unit circle and if 
it is equal to $0$, then the dynamics is along the unit circle on the
$\et_1\et_2$-plane. \figref{linear}{} illustrates the trajectories of
\eqref{linear_cc} and \eqref{linear_rr} for some choice of parameters. Each
curve shown in \figref{linear}{} corresponds to a $T_p=2\pi$-long trajectory
segment and the eigenvalues are chosen such that the qualitative differences
between two cases are easy to see.

To model the dynamics in the neighborhoods of $\PO_{1}$ and $\PO_{2}$ I
choose the latent-space dynamics given by \eqref{linear_cc} and \eqref{linear_rr},
respectively, with Floquet exponents equal to those of the periodic orbits shown
in \tabref{floquet}. As a result, the leading Floquet eigenspace of the periodic
orbits and the linear systems in \eqref{linear_cc} and \eqref{linear_rr} are
conjugates of one another and the task of the autoencoder is to perform the
coordinate transformations between these conjugate tubular neighborhoods.

\fig{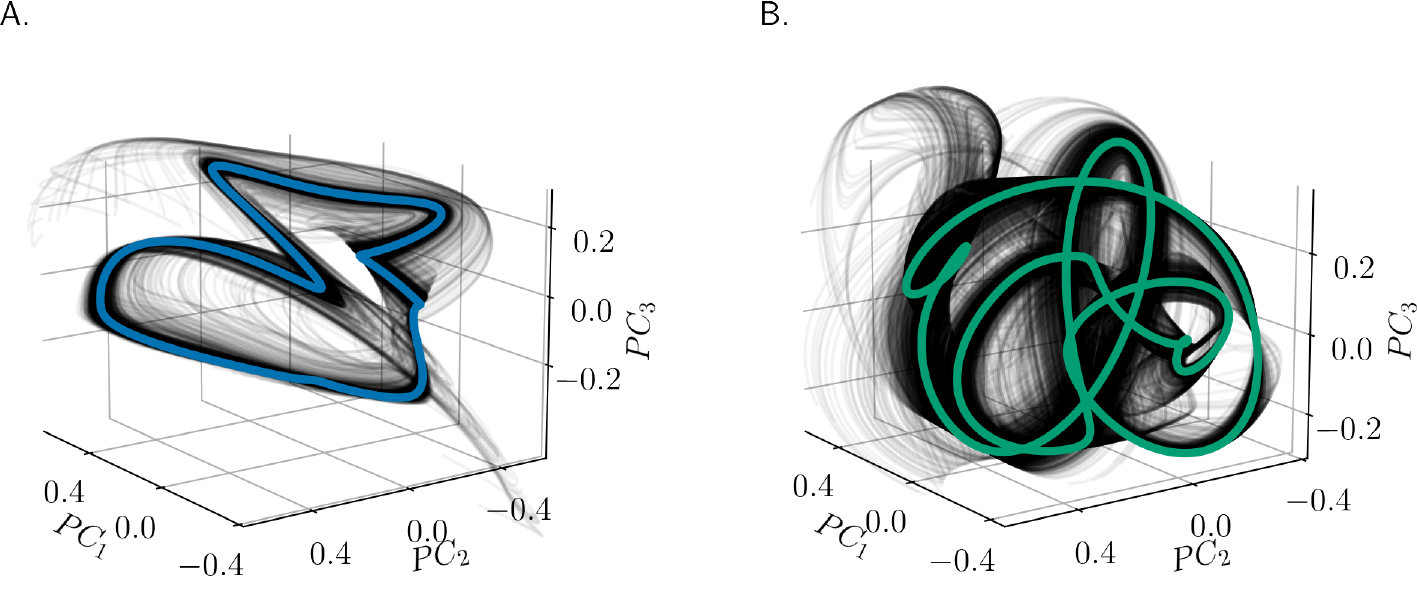}{Training data visualized as projections onto the principal
    components of the $\PO_{1}$ (A) and the $\PO_{2}$ (B).}

\subsection{Training}

Both $E$ and $D$ are multilayer perceptrons with two hidden layers, each of
which consist of $128$-nodes with \texttt{SiLU} activation functions
\cite{elfwing2018sigmoidweighted}. Together, the encoder-decoder pair have
$74785$ adjustable parameters. To produce the training data, the trajectories
initiated from $u^{(n)} (0) = u_p^{(n)} + \de u^{(n)}$, where $n = 1, 2, \ldots,
1000$ indexes the training data, $u_p^{(n)}$ is a random state on the
periodic orbit, and $\de u^{(n)}$ is a random perturbation with the amplitude
$\| \de u^{(n)} \| = 10^{-3} \| u_p^{(n)} \|$. These initial conditions were
integrated for $c / \mu_1 + T_p$, where $\mu_1$ is the real part of the
leading Floquet exponent of the periodic orbit and $T_p$ is its period.
Only the final $T_p$-long segment of these trajectories are saved
for training after symmetry reduction and the initial $c/\mu_1$-long parts
are discarded as transients. After experimenting with different values, I took
$c = 3.6$ for both periodic orbits. The training trajectories are
visualized along with the periodic orbits in \figref{trainingdata}{} as
three-dimensional projections onto the principal components of the periodic
orbits. 

After generating the data, each model is trained via Adam algorithm
\cite{kingma2014adam} to minimize a loss function. Let $\xi_p [k]$, where $k \in
\{0, 1, 2, \ldots,  \lfloor T_p / dt \rfloor \}$ be states on the periodic orbit, 
$\et_p [k]$ be the corresponding latent states on the unit circle 
and $(\xi_{l,i}, \xi_{l,f})$ be the $l$-th initial and final state pair 
sampled randomly from the $n$-th training trajectory as $\xi_{l,i} = \xi^{(n)}[0]$ and 
$\xi_{l,f} = \xi^{(n)}[m]$, where 
$n \in \{1, 2, \ldots, 1000 \} $
and $m \in \{0, 1, 2, \ldots, \lfloor T_p / dt \rfloor \} $.
The loss function is
\eqarr{
    \mathrm{Loss} &=&
            \frac{1}{\lfloor T_p / dt \rfloor + 1} \sum_{k=0}^{\lfloor T_p / dt \rfloor} 
            \Big\{ \big[ E( \xi_p [k] ) - \et_p [k] \big]^2 
            + \big[ D( E( \xi_p [k] ) ) - \xi_p [k] \big]^2 \Big\}  \nonumber \\
        &+& \frac{1}{N_b} \sum_{l = 1}^{N_b} \Big\{
            \big[ D( E( \xi_{l, i} ) ) - \xi_{l, i} \big]^2 
            + \big[ \xi_{l, f} - \xi'_{l, f} \big]^2 \Big\}
            \label{loss}
} 
where, $N_b$ is the batch size. As in \figref{architecture}, $'$ indicates
the predicted quantity, i.e. $\xi'_{l, f} = D(\oL^{m \de t} E ( \xi_{l,
i}))$. Note that only the very last term in \eqref{loss} corresponds to the
prediction errors, however, when I trained models with this error term only, I
found the transformations $E$ and $D$ to converge to a random mapping between
the training trajectories and the linear systems defined in \eqref{linear_cc}
and \eqref{linear_rr}. To remedy this, I included the first term of
the second sum in \eqref{loss}, which is the standard autoencoder loss,
penalizing the deviation of $D$ from the inverse of $E$. Finally, to ensure that
the periodic orbits are indeed mapped to the unit circle in the latent space, I
needed to include the first sum in \eqref{loss}, where the first term
penalizes the deviations of the periodic orbit from the unit circle in the
latent space and the second one is the autoencoder loss for the
periodic orbit. I initiate the training only using the periodic orbit
with a learning rate of $10^{-4}$ for $100$ epochs. Once this initial
training was complete, I provided the training data in batches of $N_b =
100$ and ran another $100$-epoch training with the learning rate $10^{-4}$ and
finally, reduced the learning rate to $10^{-5}$ and repeated. This very last
training step was terminated before $100$ epochs due to the validation losses
not improving more than $10^{-6}$ in two consecutive steps. Both models, results
from which are presented in the next section, are trained by this procedure and
the \texttt{pytorch} \cite{paszke2019pytorch} implementation using
\texttt{lightning} \cite{william2019pytorch} framework can be found in the code
repository \cite{budanur2023conjnet} accompanying this paper. 

\fig{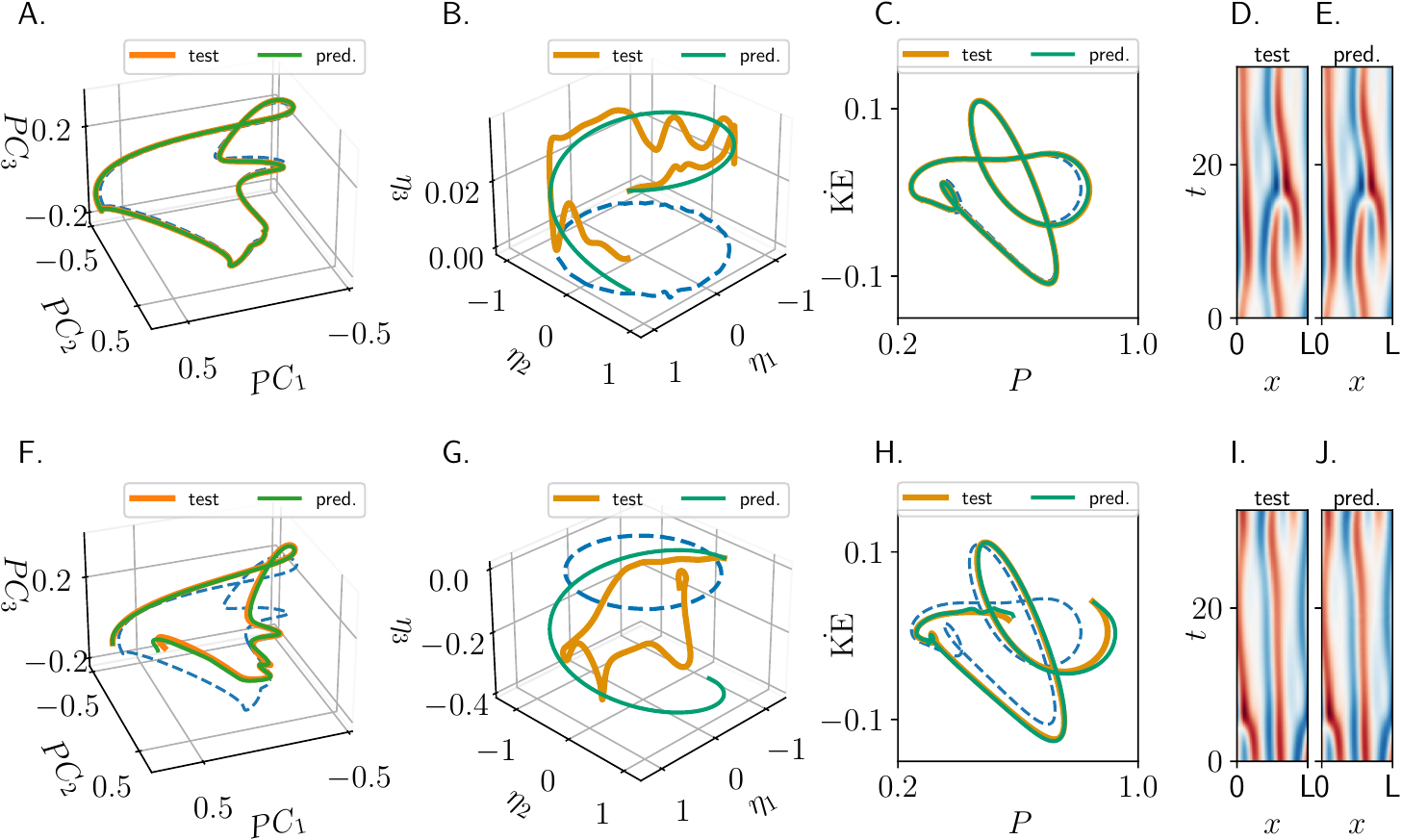}{ 
    Top (A--E) and bottom (F--J) illustrates two tests where the model predicts
    forward time evolution of the initial states in the vicinity of the periodic
    orbit $\PO_{1}$. The test trajectories (orange, thick) are visualized as
    projections onto the leading three principal components of the periodic
    orbit in the symmetry reduced state space (A and F), in the latent space (B
    and G) and on the $(P,\dot{\KE})$-plane along with the corresponding
    predictions (green) and the periodic orbit (dashed blue). (D, I) Space-time
    visualizations of the test trajectories. (E, J) Space-time visualizations of
    the predictions.  
}

\fig{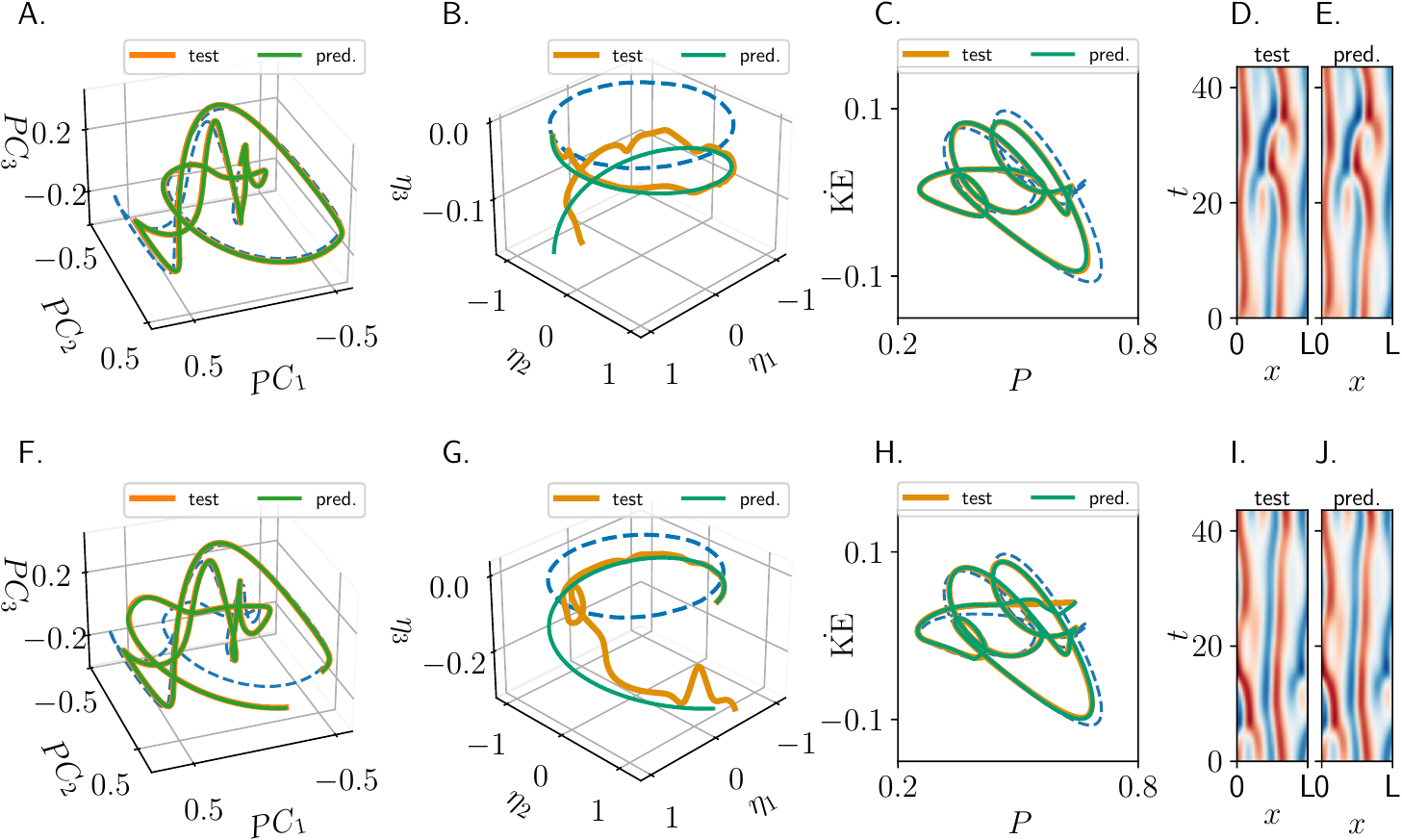}{ 
    Top (A--E) and bottom (F--J) illustrates two tests where the model predicts
    forward time evolution of the initial states in the vicinity of the periodic
    orbit $\PO_{2}$. The test trajectories (orange, thick) are visualized as
    projections onto the leading three principal components of the periodic
    orbit in the symmetry reduced state space (A and F), in the latent space (B
    and G) and on the $(P,\dot{\KE})$-plane along with the corresponding
    predictions (green) and the periodic orbit (dashed blue). (D, I) Space-time
    visualizations of the test trajectories. (E, J) Space-time visualizations of
    the predictions.  
}

\section{Results}
\label{sec/Results}

\subsection{Test trajectories}

As the first set of results, \figref{test_po1}{} shows two test trajectories,
that were initiated in the vicinity of $\PO_{1}$ by the same procedure as
the training trajectories but were not contained in the training dataset along
with the corresponding model predictions. The top panels (A--E) of
\figref{test_po1}{} show a trajectory that is initially very close to the orbit
with the minimum relative distance approximately $d_0 \approx 0.017$
and the bottom panels (F--J) illustrate another case where the initial state is
further away with $d_0 \approx 0.29$.
These relative distances were determined as 
\eq{min_rel_dist}{
    d_0 = \min_{t \in [0, T_p]} \| \xi_p(t) - \xi_0 \| / \| \xi_p(t) \|, 
    }
where $\xi_p(t)$ are the symmetry-reduced states on the periodic orbit and
$\xi_0$ is the initial symmetry-reduced state. The leftmost panels
(\figref{test_po1}{A and F}) show the test trajectories as projections onto the
$\PO_{1}$'s principal components, next to their projections in the
three-dimensional latent space (\figref{test_po1}{B and G}). \figref{test_po1}{C
and H} show the same trajectories on the plane of observables where the
horizontal and vertical axes correspond to the instantaneous rates of power
input $P$ and the time derivative of the kinetic energy $\dot{\KE} = P -
\varepsilon$, respectively. Here, I chose the $(P, \dot{KE})$-plane as opposed
to the $(P,\varepsilon)$-plane which is more frequently encountered in the
literature because the latter results in trajectories squashed around the
diagonal leaving most of the figure panel as a white space; see
\cite{cvitanovic2010on} for the comparisons of two cases. Also plotted in each
trajectory visualization \figref{test_po1}{A--C and F--H} are the corresponding
model predictions (green) and the periodic orbit (dashed blue). The rightmost
space-time plots in \figref{test_po1}{} show the symmetry-reconstructed test
trajectories and predictions, which are visually indistinguishable from one
another. The analogous test trajectories and the corresponding model predictions
for $\PO_{2}$ are shown in \figref{test_po3}{}, where the panels are organized
in the same order as \figref{test_po1}{}. These results clearly show that the
models can generalize to predict trajectories that are not included in the
training dataset. 

\fig{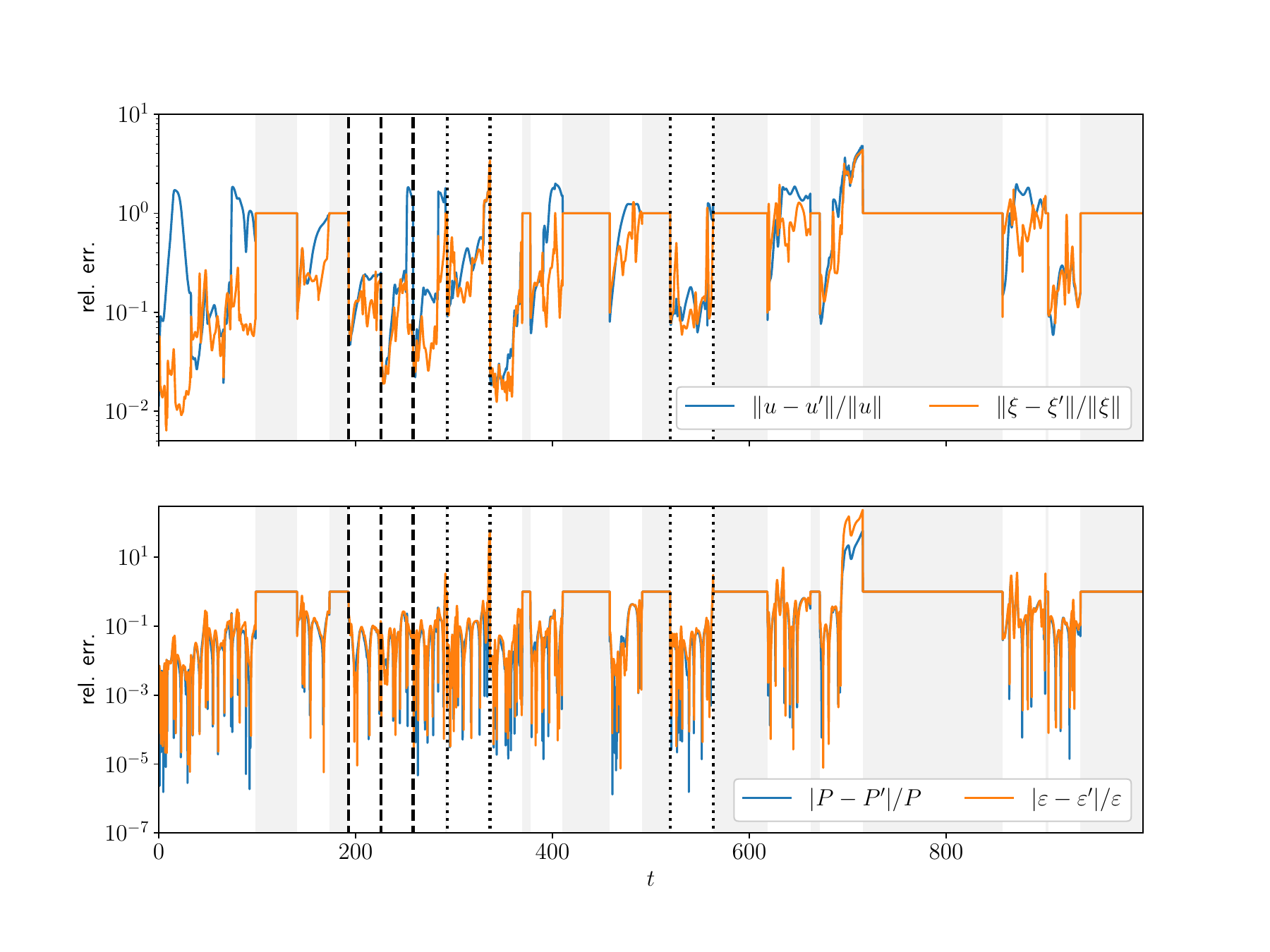}{Prediction errors in the space of the scalar fields
$u(x, t)$ and symmetry-reduced states $\xi(t)$ (top) as well as on the
domain-integrated observables power input $P$ and the rate of dissipation 
$\varepsilon$
(bottom) as a function of time. Dashed and dotted vertical lines indicate the
episodes shown in \figref{pred_po1}{} and \figref{pred_po3}{}, respectively.}

\subsection{Chaotic trajectories}
\label{sec/chaotic_trajectories}

To see whether the trained models can predict the chaotic time evolution, I
tested them using states on the trajectory depicted in \figref{spacetime}{},
treating them as a series of measurements, from which the models can
generate predictions. Because each model is intended for the neighborhood of one
of the two periodic orbits, the prediction algorithm requires a way of
determining whether a prediction can be made and, if yes, which model should be
used. This is achieved according to the following rules.
Given a state $\xi(t)$, the algorithm computes the autoencoder error 
\eq{ae_err}{
    \ep_{AE} = 
    \| \xi(t) - D ( E ( \xi(t))) \| / \| \xi(t) \| \,, 
}
for each model and finds the one with the minimum $\ep_{AE}$. If this error is
less than the threshold $\ep_{th} = 10\%$, the model is used to predict the
evolution from $t$ to $t + T_p$, where $T_p$ is the period of the associated
periodic orbit, and the algorithm repeats from $\xi(t + T_p + \de t)$. If 
the error is larger than the threshold, then the algorithm does not attempt a 
prediction and repeats from the next state $\xi (t + \de t)$.
\figref{prediction_errors}{} shows the relative errors of the predictions of the
fields $u(x,t)$ and their symmetry-reduced counterparts (top) as well as the
relative errors of the observables $P$ and $\varepsilon$ (bottom) as a function of time.
The episodes for which the $10\%$ threshold is not satisfied, thus no prediction
is made, are indicated by the gray shaded regions in \figref{prediction_errors}.
Of course, the total duration of these episodes can be reduced by
increasing the $10\%$ threshold at the expense of prediction accuracy, or
conversely, the accuracy of the predictions can be improved by reducing the
threshold. After some experimentation, I chose $10\%$ as it results in
reasonably accurate predictions as well as several inaccurate ones which I
believe are important to discuss here as examples of what can go wrong. 

\fig{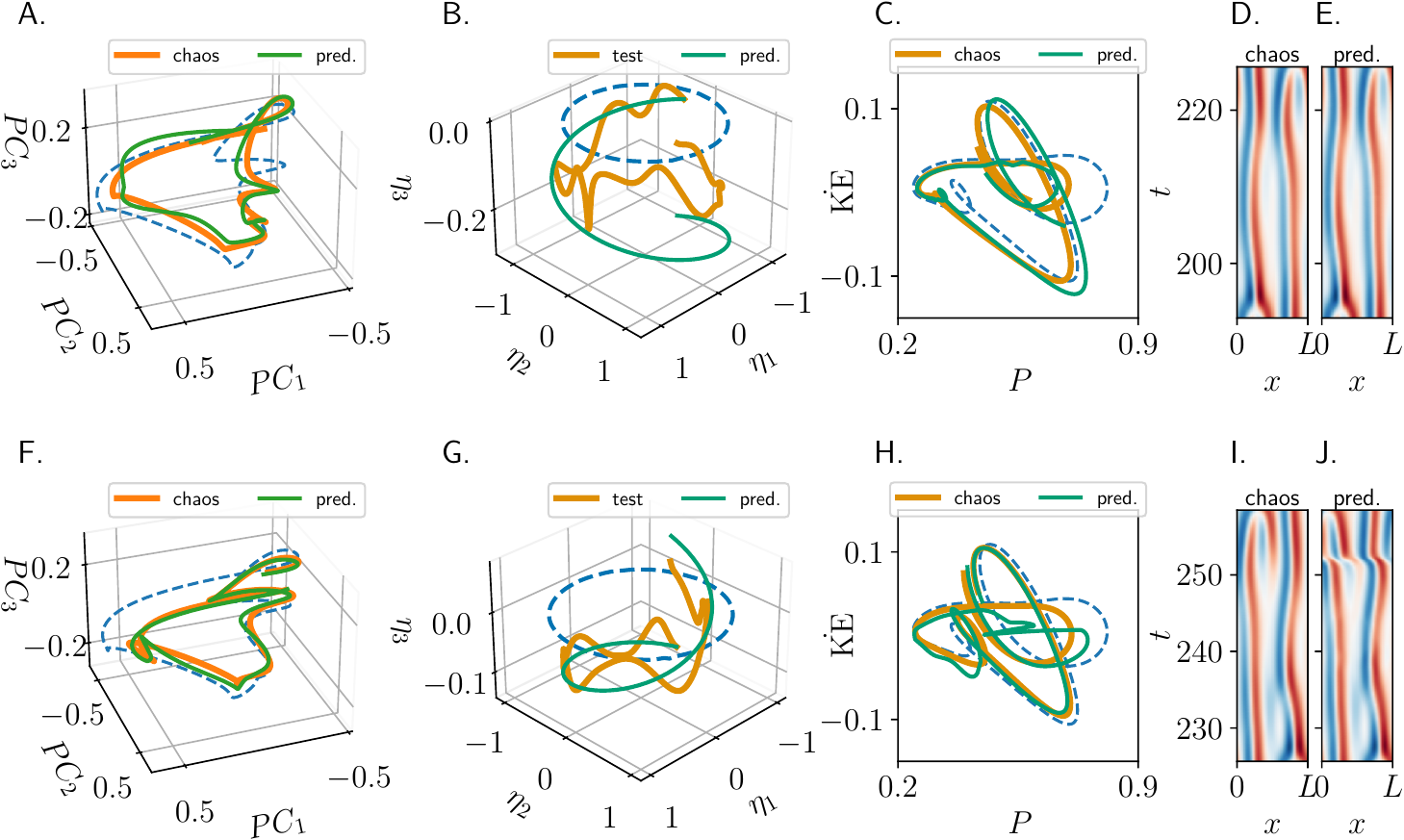}{Predictions of trajectories in the vicinity of $\PO_{1}$. Top
(A--E) and bottom (F--J) rows illustrates two different cases corresponding to
the consecutive trajectory segments along with their model predictions as PCA
projection (A, F), in the latent space (B, G), on the $(P,\dot{\KE})$-plane (C,
H), and as space-time plots (D, E, I, and J).}

\figref{pred_po1}{} illustrates two consecutive segments (top and bottom) of the
chaotic trajectory along with the corresponding predictions from the models of
$\PO_{1}$'s neighborhood. Beginning and ending of these episodes are
indicated by the vertical dashed lines in \figref{prediction_errors}{}, where
the initial time of the second episode (\figref{pred_po1}{F--J}) overlaps with
the ending of the first one (\figref{pred_po1}{A--E}). Similar to the
\figref{test_po1}{} and \figref{test_po3}{}, \figref{pred_po1}{} also shows
different visualizations of the trajectories and predictions as projections onto
the periodic orbits' principal components (A and F), in the latent space (B and
G), on the $(P, \dot{\KE})$-plane (C and H), and as space-time plots (D, E, I,
and J).  While the spacetime plots in \figref{pred_po1}{D and E} are visually
indistinguishable, \figref{pred_po1}{I and J} shows that soon after $t=250$, the
original and predicted trajectory differ significantly. Noting in
\figref{prediction_errors}{} that for this episode the relative errors in the
symmetry-reduced coordinates are still on the order of $0.1$ whereas they jump
to $1$ only for the symmetry-reconstructed predictions, one can conclude that
the error is due to the symmetry-aligning transformation. Naturally, the 
observables $P$ and $\dot{\KE}$ are not affected by this error since they are 
integrals over the domain \eqref{dEdt}, thus invariant under the 
symmetries. While the trajectories in \figref{pred_po1}{H} show 
instantaneous differences, the relative errors of the observables throughout 
this prediction window is on the order of $0.1$ as shown in 
\figref{prediction_errors}{}.
 
\fig{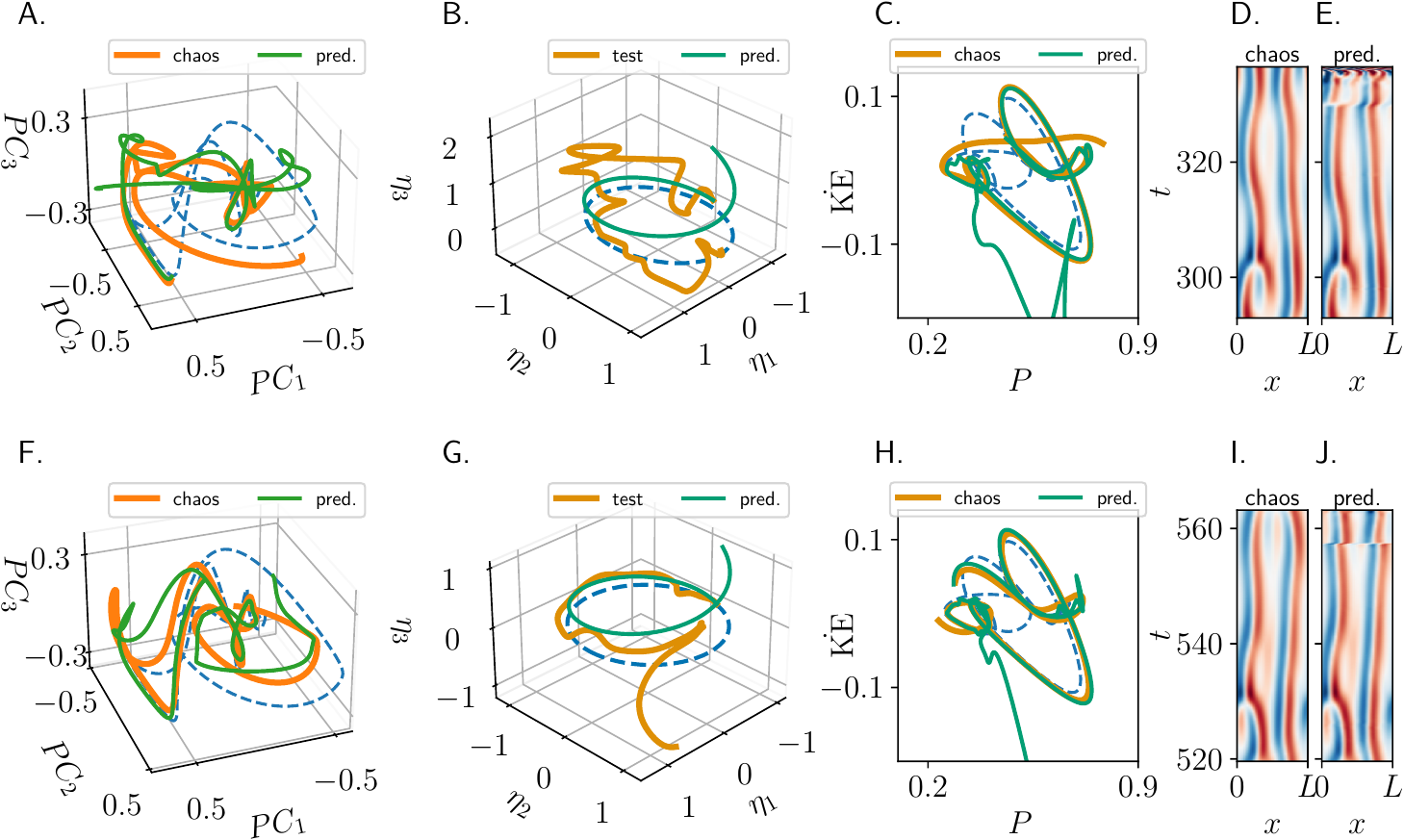}{Predictions of trajectories in the vicinity of $\PO_{2}$. Top
(A--E) and bottom (F--J) rows illustrates two different cases corresponding to
distinct trajectory segments along with their model predictions as PCA
projection (A, F), in the latent space (B, G), on the $(P,\dot{\KE})$-plane (C,
H), and as space-time plots (D, E, I, and J).}

Two trajectories in the neighborhood of $\PO_{2}$ along with the
corresponding model predictions are shown in \figref{pred_po3}{} as, once
again, projections onto the principal components of the periodic orbit (A, F),
in the latent space (B, G), on the plane of observables $(P, \dot{\KE})$ (C, H)
and as space-time plots (D, E, I, and J). The dotted vertical lines in
\figref{prediction_errors}{} indicate the beginning and endings of two episodes
that are visualized in \figref{pred_po3}{}. Notice that towards the end of the
first episode around $t \approx 330$, errors computed in both symmetry-reduced
and the reconstructed representations shoot above $1.0$. In this case, the model
diverges to a region that is outside the attractor with observable values
significantly different from those observed. This is also visible towards the
end of the space-time plot of \figref{pred_po3}{E} where the colors saturate.
Similarly, in the second case shown in \figref{pred_po3}{F--J}, the prediction
errors also significantly increase towards the end of the window as seen near
the right-most dotted vertical line in \figref{prediction_errors}{}. 

\fig{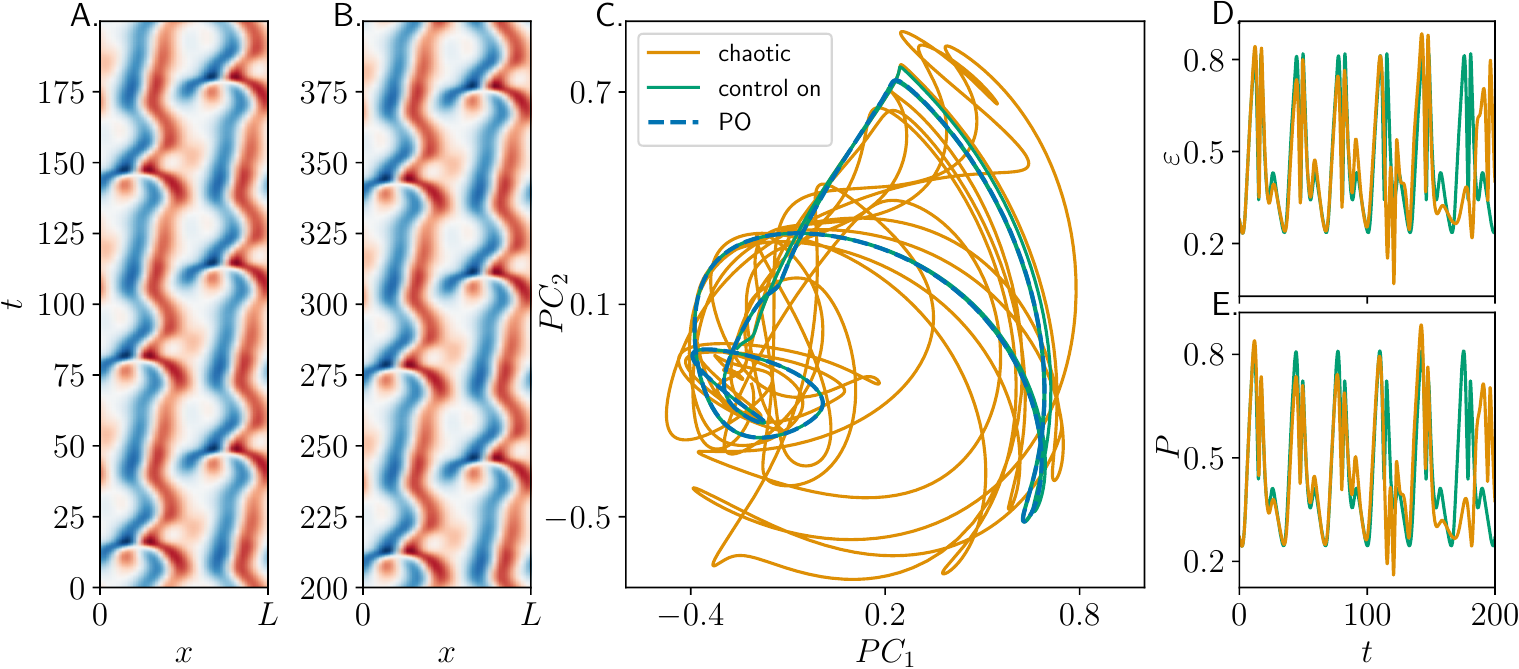}{Stabilization of the periodic orbit via control. The
space-time plots (A and B) show the periodic time-evolution of the system
starting from the same initial condition as \figref{spacetime}{} for $t \in [0,
400]$. Projection of the symmetry-reduced trajectories onto the principal
components of the periodic orbit (C): Chaotic trajectory for $t \in [0, 200]$
(orange), Controlled trajectory for $t \in [0, 200]$ (green), and the periodic
orbit $\PO_{1}$ (dashed blue). Time series of the instantaneous rate of
dissipation ($\varepsilon$) and power input ($P$) of the chaotic (orange)
solution and the solution of the controlled system (green).}

\subsection{Control}

To demonstrate how the understanding the latent-space dynamics of the presented
models can be leveraged for applications, I would like to present a simple
control method that stabilizes $\PO_{1}$. Let $F(u)$ denote the RHS of the
Kuramoto--Sivashsinky equation \eqref{KS}, i.e. $u_t=F(u)$, and consider the
controlled system 
\eq{controlled_KS}{u_t = F(u) + H(u)} where 
\eq{control}{
H(u) = 
    \begin{cases}
            0 & \text{if } \ep_{AE} > \ep_{th} \,, \\
            h \ga^{*} S^{-1} (D (\et / \| \et \| - \et)) & \text{otherwise.} 
    \end{cases}
}
In the equation above, $h \in \reals_{>0}$ is a constant ``gain'', $\ga^{*} =
\argmin_{\ga \in \GA} \| u - \ga S^{-1}( S ( u) ) \|$ is the state-aligning
symmetry operator, and $\et = E (S (u))$ is the encoded state. Notice that the
term in the innermost parentheses in \eqref{control} penalizes the latent
states' norm's difference from $1$ and becomes $0$, if $\| \et \| = 1$. 

\figref{stabilizing_control}{A and B} show the space-time plot of the $u(x,t)$
obtained from the simulation of \eqref{controlled_KS} with $h = 0.4$ and
$\ep_{th} = 0.1$ starting from the same initial condition as
\figref{spacetime}{}. As seen, the dynamics has now become relative periodic as
opposed to the chaotic shown in \figref{spacetime}{}. For further illustration,
\figref{stabilizing_control}{C} shows the symmetry-reduced controlled trajectory
(green) along with the chaotic trajectory segment (orange) for $t \in [0, 200]$
as a projection onto the principal components of the $\PO_{1}$, which itself
is shown as the dashed blue curve. Notice that the stabilized green trajectory
and the dashed blue one mostly overlap in \figref{stabilizing_control}{C}.
Finally, the time periodicity can also be seen in the time-series of the
observables $\varepsilon$ and $P$, which are plotted green in
\figref{stabilizing_control}{D and E}, respectively, along with the
corresponding time-series obtained from the chaotic trajectory in orange. 

\section{Discussion}
\label{sec/Discussion}

With the results shown in \figref{pred_po1}{} and \figref{pred_po3}{}, I
attempted to demonstrate that the chaotic trajectories of the
Kuramoto--Sivashinsky system can be predicted using models with linear
latent-space dynamics. Among the four cases presented, three of them have
visible discrepancies towards the end of the prediction window, which I would
like to comment on and speculate possible remedies. 

First, I would like to focus on \figref{pred_po3}{D and E} where at the very end
of the prediction window the original and the predicted trajectories  
differ significantly.
This is due to the rapid growth of the predictions' amplitude, which is also
manifested in \figref{pred_po3}{C} by the final segment's exit from the area
shown. Presumably, the very last segment of the corresponding latent-space
trajectory lays in a region which was not included in the training data, which
in turn results in its decoding into a trajectory outside the attractor. A
similar divergence also takes place at the very end of the second case shown in
the bottom row of \figref{pred_po3}{}.  In practice, one could discard such
unphysical predictions by comparing their amplitude to the attractor limits.
Another way of eliminating divergences could be further training the neural
networks using data collected from the attractor or fine-tuning the transient
times at the training stage. 

Another discrepancy that is best seen in \figref{pred_po1}{J} and
\figref{pred_po3}{J} are predicted peaks appearing in locations that are shifted
in space, or mirrored locations with respect to the reference solutions. These
issues are due to the difficulties with reconstructing discrete-symmetry-reduced
trajectories, which relies on the numerical estimation of the partial
derivative $u'_t$. As stated in \secref{basic_assumptions}, this
is a source of numerical error since it is obtained via finite differences. In
my initial tests of the symmetry aligning transformations by applying them to
the symmetry-reduced trajectories $\xi(t)$, errors such as those seen in
\figref{pred_po1}{J} and \figref{pred_po3}{J} were not present. However, when 
trajectories were obtained by models via decoding the linearly-evolved 
latent-space curves, I observed momentary jumps in the decoded trajectories, which 
in turn resulted in selection of false symmetry copies.
I would like to emphasize that if one is interested in predicting the
domain-integrated observables, then this is no issue, since such observables are
symmetry-invariant. 
For discrete symmetry reduction, I produced
symmetry-reducing complex variables through an optimization over the attractor.
In future iterations of the modeling approach presented here, I plan to explore
carrying out such optimizations also locally in the state space, since the
models that are local in the state space could feasibly employ symmetry-reducing
transformations that are also local. Another idea would be abandoning discrete
symmetry reduction altogether at the expense of having pre-periodic orbits become
periodic only after two periods. This, however, would potentially introduce
performance issues due to the exponential divergence of trajectories. 

While I chose to present the prediction results with forecast windows equal to
the periods of the orbits, I would like to note that this was a pedagogical
choice rather than a necessity. One straightforward way of reducing the
prediction errors is indeed would be reducing the prediction windows. In units
of Lyapunov time $t_{L} \approx 20.9$, which I estimated via the Benettin
algorithm \cite{benettin1976kolmogorov}, the periods of $\PO_{1}$ and
$\PO_{2}$ are approximately $1.6 t_L$ and $2.1 t_L$, respectively. It can be
readily seen from \figref{prediction_errors}{} that reducing the prediction
window to, for instance, one Lyapunov time could eliminate the episodes with
larger errors towards the end of the prediction windows indicated by the dashed
and dotted lines.

\section{Conclusion \& Outlook}
\label{sec/Conclusion}

In this paper, I presented a reduced-order modelling tool that combines the
knowledge of numerically exact periodic orbits with the neural networks to
produce predictive models of spatiotemporal chaos in the Kuramoto--Sivashinsky
system. My primary goal was to demonstrate that such models can be significantly
simple and exhibit some degree of interpretability if they are built around the
periodic orbits. To illustrate the usefulness of understanding models, I
implemented an orbit stabilization method wherein the control input was
determined in the latent space. 

Although Kuramoto--Sivashinsky system shares several similarities with the
moderate-Reynolds-number turbulent flows with which I motivated this study,
there are several challenges that need to be addressed when applying the
tools presented here to such problems. The first of these is that moving
from one spatial dimension to two- and three- leads to a substantial
increase in the number of degrees of freedom as it grows geometrically with
the spatial dimension. When this becomes prohibitive for neural network
tools, one could include a PCA projection between the symmetry reduction and
encoder in \figref{architecture}{} as an initial dimensionality reduction
step to ease the computational burden. While the results presented here
showed that the two periodic orbits were sufficient to capture nearly $50\%$
of the time evolution, it would be too optimistic to expect such a simple
phenomenology in higher-dimensional settings. Thus, when moving towards two-
and three-dimensional fluid flows, one should be ready to face the
difficulties of finding many periodic orbits in higher-dimensional settings.
Such future efforts could presumably benefit from the adjoint-based periodic
orbit search tools \cite{azimi2022constructing,parker2022variational} which
typically exhibit better convergence behavior in comparison to the
Newton-based methods.

An obvious difficulty of the present method in comparison to the other
data-driven model reduction tools is the need for finding periodic orbits, which
renders it impossible to apply to observational data collected from systems
without detailed models. One potential way of eliminating this need could be
utilizing DMD to approximate nearly periodic dynamics borrowing ideas from Refs.
\cite{page2020searching,marensi2023symmetryreduced}, where it was shown that DMD
can be used to approximate unstable (relative) periodic orbits in shear flow
simulations. Another way of eliminating the dependence on detailed models could
be developing clustering methods that incorporate dynamics by considering
bundles of trajectories as opposed to states sampled on an attractor. These are
the research directions which I hope to explore in the future. 

\appendix
\section{Symmetry reduction of the Kuramoto--Sivashinsky system}
\label{app/symmred}

In Fourier space, the translations $\oT_{\de x} u(x,t) = u(x - \de x,t)$ 
correspond to the rotations $\oR (\th) v_k = e^{i k \th} v_k$ where 
$\th = 2 \pi \de x / L$. Let $\ph_2 (t) = \angle v_2 (t)$ be the complex phase of the 
second Fourier mode, it is straightforward to confirm that the transformation 
\eq{phase_fixing}{
    v'_k = \ga \left(\frac{\pi}{4} - \frac{\ph_2}{2} \right) v_k 
}
sets the phase of the $v'_k$ to $\pi/2$. Although its exact value is irrelevant
at this stage, setting this to $\pi/2$ is going to be convenient later. Unlike fixing 
the phase of the first Fourier mode, however, the transformation \eqref{phase_fixing}
does not fully eliminate the translations, since  $\oR  (\pi) v'_2 = v_2$ also 
satisfies the phase condition, whereas the odd modes pick up a phase of $e^{i \pi}$.
In other words, the reduced dynamics in the space of $v'_k$ have a  $\pi$-rotation 
symmetry which can be expressed as 
\eq{pi_symm}{
    \oR (\pi) v'_k = 
    \begin{cases}
        e^{i \pi} v'_k &\quad \text{if k is odd} \,, \\
        v'_k &\quad \text{if k is even} \,. 
    \end{cases}
}
This multiplicity is easily eliminated by the 2-to-1 transformation 
\eq{pi_symm_red}{
    v''_k = 
    \begin{cases}
        e^{i \ph'_1} v'_k &\quad \text{if k is odd} \,, \\
        v'_k &\quad \text{if k is even} \,, 
    \end{cases}
}
where $\ph'_1 = \angle v'_1$. Noting that $\ph'_1 \rightarrow \ph'_1 + \pi$
under the action of $\oR  (\pi)$, one can confirm that $v''_k$ are invariant
under this symmetry. The discrete symmetry reducing transformation
\eqref{pi_symm_red} is a simple application of the general method for reducing
cyclic symmetries which will be presented in a separate publication
\cite{kneer2023learning}. The final remaining symmetry of the system is the
reflection and in order to reduce it, one first needs to find its action on the
translation-invariant modes $v''_k$. As stated earlier, the phase condition
$\ph'_2 = \pi/2$ which ensures that $v'_2$ is purely imaginary is convenient
since it is invariant under the action of reflection $\oM v_k = -v_k^*$, where
$^*$ indicates complex conjugation. Thus, the action of $\oM$ on the modes
$v'_k$ is the same as that on $v_k$. The $\pi$-reducing transformation
\eqref{pi_symm_red}, however, introduces an additional phase factor and which
yields the action of $\oM$ on $v''_k$ as 
\eq{pi_symm}{
    \oM v''_k = 
    \begin{cases}
        (v''_k)^* &\quad \text{if k is odd} \,, \\
        - (v''_k)^* &\quad \text{if k is even} \,. 
    \end{cases}
}
In order to follow the same recipe as the reduction of the rotation-by-$\pi$
symmetry, one first needs to express the symmetry action as a complex phase 
$e^{i \pi}$. This is achieved grouping the symmetry-invariant parts 
$\Re v''_{1,3,5,\ldots}$ and $\Im v''_{0, 2, 4, \ldots}$ 
and sign changing components
$\Im v''_{1,3,5,\ldots}$ and $\Re v''_{0, 2, 4, \ldots}$ into new variables 
$w^+_k$ and $w^-_k$ such that 
$\oM w^+_k = w^+_k$ and $\oM w^-_k = e^{i \pi} w^-_k$. Finally, one can generate 
a symmetry-reducing variable 
\eq{what}{
    \hat{w} = \sum_k c_k w^-_k \,, 
}
where $c_k$ are coefficients and use its phase $\ph_\oM = \angle \hat{w}$ to
transform $w^-_k$ as $\hat{w}^-_k = e^{i \ph_\oM} \hat{w}^-_k$, which are
invariant under $\oM$. For the results of this paper, I determined $c_k$ by
maximizing the amplitude of $\hat{w}$ \eqref{what} in order to avoid fast phase
oscillations. These coefficients along with the \texttt{python} implementations
of forward and backward transformations are openly available in the code
repository \cite{budanur2023conjnet} accompanying this paper. 

\bibliography{conj}

\end{document}